\documentclass[12pt,preprint]{aastex}
\usepackage{longtable}

\begin{document}

\title{Cyclic period oscillation of the eclipsing dwarf nova DV UMa}

\author{Han Z.-T\altaffilmark{1,2,3}, Qian S.-B\altaffilmark{1,2,3}, Irina Voloshina\altaffilmark{4}, Zhu L.-Y\altaffilmark{1,2,3}}

\singlespace

\altaffiltext{1}{Yunnan Observatories, Chinese Academy of Sciences (CAS), P. O. Box 110, 650216 Kunming, China; zhongtaohan@ynao.ac.cn}
\altaffiltext{2}{Key Laboratory of the Structure and Evolution of Celestial Objects, Chinese Academy of Sciences, P. O. Box 110, 650216 Kunming, China}
\altaffiltext{3}{University of Chinese Academy of Sciences, Yuquan Road 19\#, Sijingshang Block, 100049 Beijing, China}
\altaffiltext{4}{Sternberg Astronomical Institute, Moscow State University, Universitetskij prospect 13, Moscow 119992, Russia}

\begin{abstract}
DV UMa is an eclipsing dwarf nova with an orbital period of $\sim2.06$ h, which lies just at the bottom edge of the period gap.
To detect its orbital period changes we present 12 new mid-eclipse times by using our CCD photometric data and archival data.
Combining with the published mid-eclipse times in quiescence, spanning $\sim30$ yr, the latest version of the $O-C$ diagram was obtained and analyzed.
The best fit to those available eclipse timings shows that the orbital period of DV UMa is undergoing a cyclic oscillation with a period of
$17.58(\pm0.52)$ yr and an amplitude of $71.1(\pm6.7)$ s. The periodic variation most likely arises from the light-travel-time effect via the presence of
a circumbinary object because the required energy to drive the Applegate mechanism is too high in this system. The mass of the unseen companion was derived as $M_{3}\sin{i'}=0.025(\pm0.004)M_{\odot}$.
If the third body is in the orbital plane (i.e. $i'=i=82.9^{\circ}$) of the eclipsing pair, it would match to a brown dwarf.
This hypothetical brown dwarf is orbiting its host star at a separation of $\sim8.6$ AU in an eccentric orbit ($e=0.44$).
\end{abstract}


\section{Introduction}
Cataclysmic variables (CVs) are short-period binaries containing a white dwarf and a low-mass donor star that is transferring material to the white dwarf via an accretion disc (Warner 1995).
The structure of CVs allows precise timing measurements because the components have large differences in radius and luminosity (e.g. Parsons et al. 2010).
The timing measurements offer important clues concerning the long-term evolution of orbital periods and the existence of circumbinary substellar objects.
By analyzing the observed$-$calculated ($O-C$) curve of these systems, the orbital period and its rate of change can be measured. The secular change of the $O-C$ curve can provide key information on the evolution of CVs (Qian et al. 2015). Moreover, if there is a third body orbiting the close binary system, it will cause a small wobble cyclically in the timing of eclipses.
More specifically, if the presence of a third body, then the binary system and companion revolve around their common barycenter.
The observed times of eclipses while the system is on the near side of the larger orbit will be detected sooner than on the far side. This will lead to alternating variations of the observed eclipse timings, which is often referred to as the light$-$travel$-$time (LTT) effect.
The LTT shows a periodic variation in $O-C$ diagram.
The timing method has recently been used to detect possible extrasolar planets around CVs such as V2051 Oph (Qian et al. 2015), Z Cha (Dai et al. 2009), OY Car (Han et al. 2015) and V893 Sco (Bruch 2014).

As a member of the dwarf nova-type CVs, DV UMa was first discovered as an ultraviolet excess object by Usher et al. (1982). Subsequently, this star was identified as a candidate dwarf nova based on the observed outburst phenomena and the $H\alpha$ emission line in optical spectra (Usher et al. 1983). Howell et al. (1987) presented photometric
observations, which showed large brightness changes on a shorter time-scale. Further photometric observations by Howell et al. (1988) revealed that DV UMa is an eclipsing system, and its orbital period was estimated as $0.08579(1)$ d. The system parameters were derived by Howell \& Blanton (1993) using photometric
analysis and by Szkody \& Howell (1993) using time-resolved spectroscopy. After that, these parameters have been improved to give higher precision (e.g. Patterson et al. 2000, Feline et al. 2004 and Savoury et al. 2011). However, we still know little about its evolution and period changes. Although mid-eclipse times of DV UMa have been published in the
literature and the orbital ephemeris has been updated by several authors, no sign of orbital period change was found (Howell et al. 1988, Patterson et al. 2000, Nogami et al. 2001, Feline et al. 2004). In the present paper, we
present new CCD photometric observations of DV UMa
and detect a cyclic variation in the $O-C$ diagram.
Then the presence of a substellar companion is discussed.

\section{Observations and data preparation}

New CCD photometric observations of DV UMa were carried out by using three different telescopes.
Beginning on 2009 November 12, this star was continuously monitored with the 2.4$-$m telescope at the Lijiang observational station of Yunnan
Observatories (YNAO).
To get more data for this binary, an observation was made on 29 March, 2016 with an Andor DW436 1K CCD camera mounted on
the 85$-$cm reflecting telescope at the XingLong station of the National Astronomical Observatory, and another observation made on 23 January, 2017 with an Andor DW936 2K CCD camera attached to the 1.0$-$m reflecting telescope at YNAO. During the observations,
no filters were used, in order to improve the time
resolution. All observed CCD images were analyzed
by applying the aperture photometry package of IRAF.
Differential photometry was performed, with a nearby
non-variable comparison star. Four eclipse profiles observed with the 2.4$-$m and 1.0$-$m are displayed in Fig. 1.
To measure the mid$-$eclipse times ($T_{mid}$) of the white dwarf we use the method described by Wood et al. (1985).
First, the mid$-$ingress ($T_i$) and mid$-$egress times ($T_e$) of the white dwarf were determined by locating the maximum and minimum values of the derivative of the light curves. Then the mid$-$eclipse times were derived by using $T_{mid}=(T_i+T_e)/2$.
The procedure of measuring $T_{mid}$ is illustrated in Fig. 2. It is important to note that only data during quiescence were used to determine
the mid$-$eclipse times, because the system in outburst is dominated by the accretion disc, and the $T_i$ and $T_e$ of the white dwarf are not clearly identified in the derivative curve.
But during quiescence the ingress and egress times of the white dwarf are stable features.
As shown in Fig. 2, the white dwarf and bright-spot ingress and egress are both clear and distinct.
With our data, eight mid$-$eclipse times were obtained. The errors are the standard errors
in measuring mid-eclipse times, and they depend on the time resolution and
signal-to-noise ratio during observations.
Apart from our data, four mid-eclipse times during quiescence were also determined with the observations from American Association of Variable Star Observers (AAVSO).
By checking the AAVSO database we find that their observations contain many eclipsing light curves. Moreover, the long-term AAVSO data will help to confirm whether this star is in outburst or not. These new mid$-$eclipse times and their errors were listed in Table 1.

\begin{figure}[!h]
\begin{center}
\includegraphics[width=16cm]{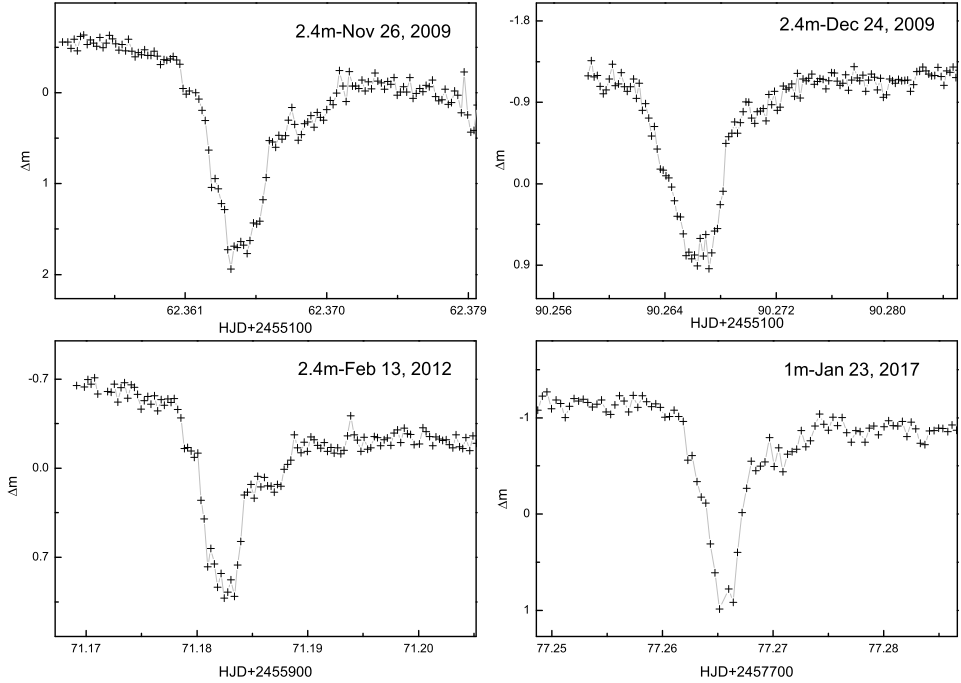}
\caption{Four eclipse profiles of DV UMa in quiescence obtained at YNAO using the 1.0m and 2.4m telescopes.}
\end{center}
\end{figure}

\begin{figure}[!h]
\begin{center}
\includegraphics[width=16cm]{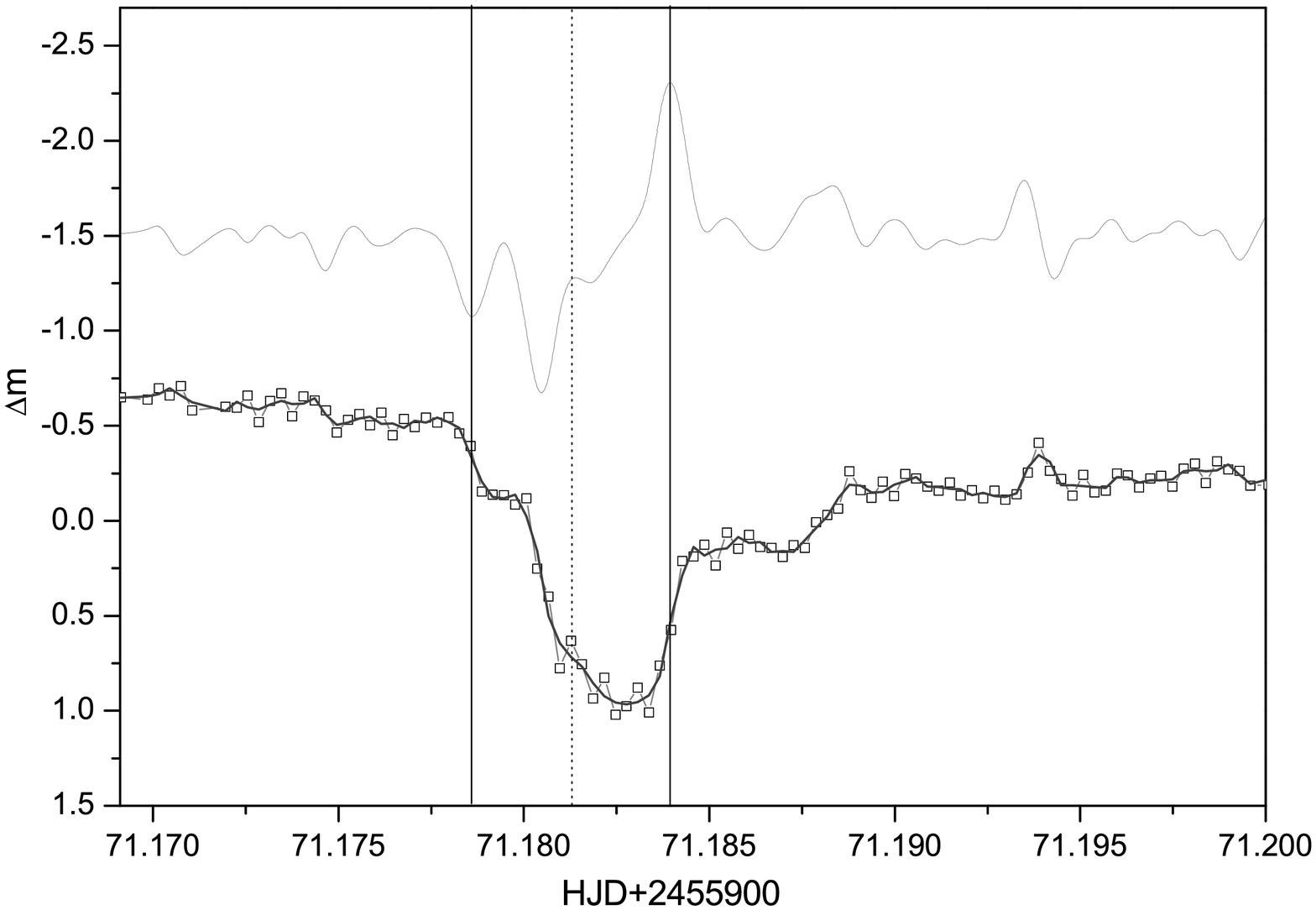}
\caption{A sample of determining mid-egress times. The boxes denote the observed light curve, and the solid curve is the corresponding smoothed light curve. The derivative of the smoothed curve with a spline fit is plotted in the upper part of the diagram. Two vertical solid lines refer to the mid-ingress and mid-egress times of the white dwarf, while the dashed vertical line refers to the mid-eclipse times.}
\end{center}
\end{figure}

Mid-eclipse times of DV UMa also have been published in the literatures by several previous authors. Howell et al. (1988) first reported twelve mid$-$eclipse times and given an orbital ephemeris.
Later, Patterson et al. (2000) updated the orbital ephemeris by adding twelve mid$-$eclipse times. After just one year, the ephemeris was revised again by Nogami et al. (2001).
A recent version of the orbital ephemeris was determined by Feline et al. (2004).
However, not all published data are suitable for the period analyses. For instance, some mid$-$eclipse times during outbursts should be excluded.
Based on the observations of the outbursts reported by Patterson et al. (2000) and Nogami et al. (2001), the mid$-$eclipse times during outbursts can be separated from these historical data.
These data and related information are also given in Table 1.
The linear ephemeris presented by Nogami et al. (2001) is used to compute the $O-C$ values of all mid$-$eclipse times, which are plotted in Fig. 3.

\begin{figure}[!h]
\begin{center}
\includegraphics[width=16cm]{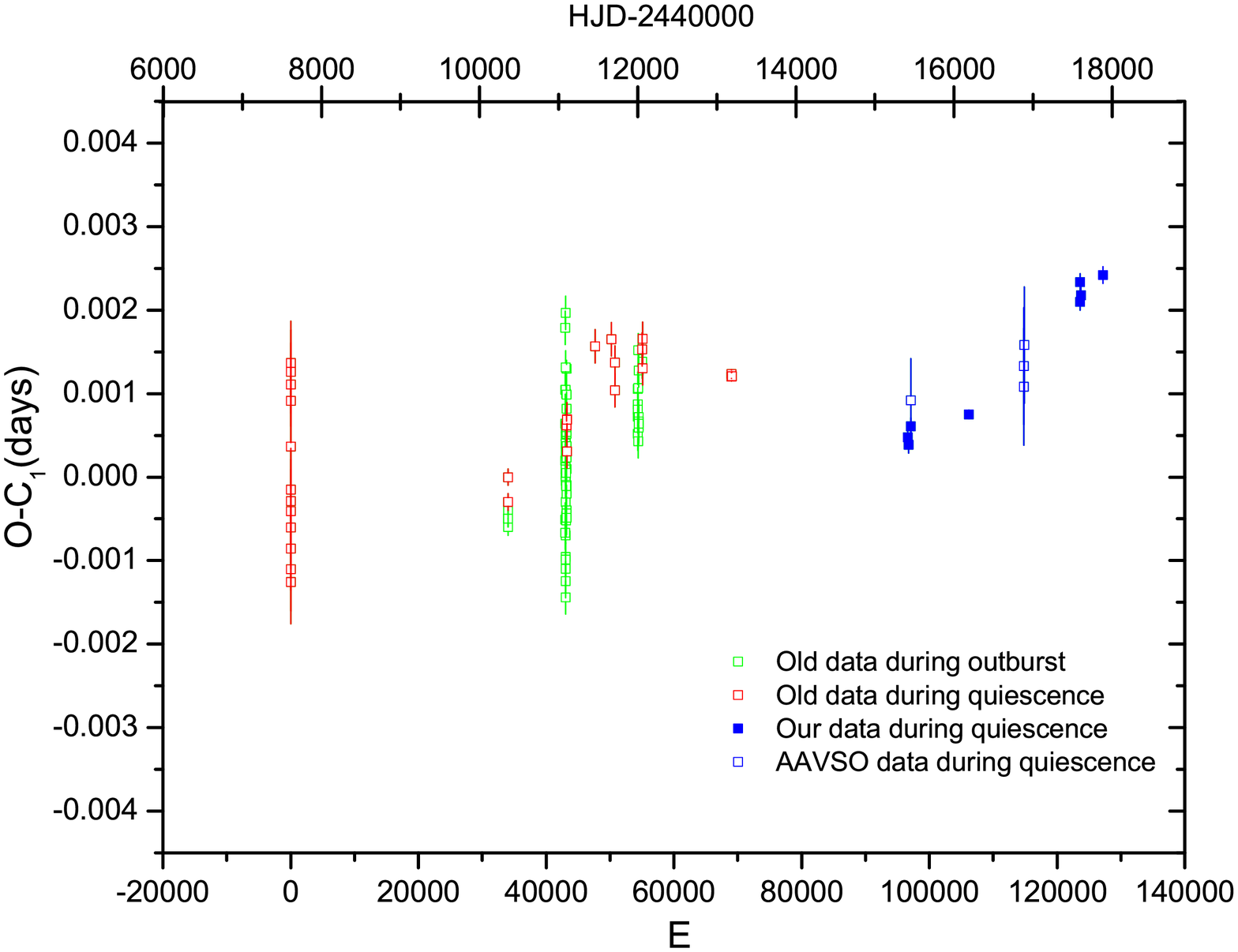}
\caption{The latest $O-C$ diagram of DV UMa, using new mid-eclipse times together with all published data. New eclipse timings in quiescence are denoted by the blue boxes (open and solid). Those data obtained from the literature can be divided into two groups: during outburst marked by the green boxes and during quiescence depicted by the red boxes.}
\end{center}
\end{figure}

\begin{table*}
\caption{All mid-eclipse times of DV UMa.}
 \begin{center}
 \small
 \begin{tabular}{ccccccc}\hline\hline
Min.(HJD)     &E       &O-C        &Errors   & State       &Telescopes      &Ref.$^a$    \\\hline
2446854.7485  &1	     &-0.00015   &0.00050  & quiescence   &   $-$         &  (1)    \\
2446854.8339  &2	     &-0.00060   &0.00050  & quiescence   &   $-$         &  (1)    \\
2446854.9195  &3	     &-0.00086   &0.00050  & quiescence   &   $-$         &  (1)    \\
2446855.6944  &12      &0.00137    &0.00050  & quiescence   &   $-$         &  (1)    \\
2446855.7798  &13      &0.00092    &0.00050  & quiescence   &   $-$         &  (1)    \\
2446855.8660  &14      &0.00126    &0.00050  & quiescence   &   $-$         &  (1)    \\
2446855.9517  &15      &0.00111    &0.00050  & quiescence   &   $-$         &  (1)    \\
2446856.4646  &21      &-0.0011    &0.00050  & quiescence   &   $-$         &  (1)    \\
2446856.5503  &22      &-0.00126   &0.00050  & quiescence   &   $-$         &  (1)    \\
2446856.6370  &23      &-0.00041   &0.00050  & quiescence   &   $-$         &  (1)    \\
2446857.4963  &33      &0.00036    &0.00050  & quiescence   &   $-$         &  (1)    \\
2446857.5815  &34      &-0.00029   &0.00050  & quiescence   &   $-$         &  (1)    \\
2449775.9693  &34027   &-0.00050   &0.00020  & outburst     &   $-$         &  (2)    \\
2449776.0552  &34028   &-0.00040   &0.00020  & outburst     &   $-$         &  (2)    \\
2449776.1409  &34029   &-0.00060   &0.00020  & outburst     &   $-$         &  (2)    \\
2449776.2269  &34030   &-0.00050   &0.00020  & outburst     &   $-$         &  (2)    \\
2449778.5451  &34057   &-0.00030   &0.00020  & quiescence   &   $-$         &  (2)    \\
2449778.6312  &34058   &0.00000    &0.00020  & quiescence   &   $-$         &  (2)    \\
2450548.3860  &43024   &0.00020    &0.00010  & outburst     &   $-$         &  (3)    \\
2450548.4727  &43025   &0.00104    &0.00010  & outburst     &   $-$         &  (3)    \\
2450548.5570  &43026   &-0.00051   &0.00010  & outburst     &   $-$         &  (3)    \\
2450548.6440  &43027   &0.000639   &0.00010  & outburst     &   $-$         &  (3)    \\
2450548.7310  &43028   &0.00179    &0.00010  & outburst     &   $-$         &  (3)    \\
2450548.8144  &43029   &-0.00067   &0.00010  & outburst     &   $-$         &  (3)    \\
2450549.0730  &43032   &0.00040    &0.00020  & outburst     &   $-$         &  (2)    \\
2450549.1578  &43033   &-0.00070   &0.00020  & outburst     &   $-$         &  (2)    \\
2450549.4180  &43036   &0.00197    &0.00010  & outburst     &   $-$         &  (3)    \\
2450549.5032  &43037   &0.00131    &0.00010  & outburst     &   $-$         &  (3)    \\
2450550.3609  &43047   &0.00049    &0.00010  & outburst     &   $-$         &  (3)    \\
2450550.4463  &43048   &0.00003    &0.00010  & outburst     &   $-$         &  (3)    \\
2450551.3894  &43059   &-0.00124   &0.00010  & outburst     &   $-$         &  (3)    \\
2450551.4754  &43060   &-0.00110   &0.00010  & outburst     &   $-$         &  (3)    \\
2450551.7351  &43063   &0.00105    &0.00010  & outburst     &   $-$         &  (3)    \\
2450551.8202  &43064   &0.00029    &0.00010  & outburst     &   $-$         &  (3)    \\
2450551.9916  &43066   &0.00000    &0.00020  & outburst     &   $-$         &  (2)    \\
2450552.0776  &43067   &0.00010    &0.00020  & outburst     &   $-$         &  (2)    \\
2450552.1630  &43068   &-0.00030   &0.00020  & outburst     &   $-$         &  (2)    \\
2450552.4214  &43071   &0.00050    &0.00020  & outburst     &   $-$         &  (2)    \\
2450552.5927  &43073   &0.00012    &0.00010  & outburst     &   $-$         &  (3)    \\
2450553.3643  &43082   &-0.00095   &0.00010  & outburst     &   $-$         &  (3)    \\
2450553.4512  &43083   &0.00009    &0.00010  & outburst     &   $-$         &  (3)    \\
2450553.6234  &43085   &0.00059    &0.00010  & outburst     &   $-$         &  (3)    \\
2450553.7093  &43086   &0.00064    &0.00010  & outburst     &   $-$         &  (3)    \\
2450554.3945  &43094   &-0.00099   &0.00010  & outburst     &   $-$         &  (3)    \\
2450554.4799  &43095   &-0.00144   &0.00010  & outburst     &   $-$         &  (3)    \\\hline
\end{tabular}
\end{center}
\end{table*}

\addtocounter{table}{-1}
\begin{table*}
\caption{$-$continued.}
 \begin{center}
 \small
 \begin{tabular}{ccccccc}\hline\hline
Min.(HJD)     &E       &O-C        &Errors   & State       &Telescopes      &Ref.$^a$    \\\hline
2450554.5672  &43096   &0.00001    &0.00010  & outburst     &   $-$         &  (3)    \\
2450554.6530  &43097   &-0.00004   &0.00010  & outburst     &   $-$         &  (3)    \\
2450554.7393  &43098   &0.00040    &0.00010  & outburst     &   $-$         &  (3)    \\
2450554.8248  &43099   &0.00005    &0.00010  & outburst     &   $-$         &  (3)    \\
2450555.3394  &43105   &-0.00046   &0.00010  & outburst     &   $-$         &  (3)    \\
2450555.4252  &43106   &-0.00052   &0.00010  & outburst     &   $-$         &  (3)    \\
2450555.4267  &43106   &0.00100    &0.00020  & outburst     &   $-$         &  (2)    \\
2450557.6584  &43132   &0.00051    &0.00010  & outburst     &   $-$         &  (3)    \\
2450557.7441  &43133   &0.00036    &0.00010  & outburst     &   $-$         &  (3)    \\
2450557.8295  &43134   &-0.00009   &0.00010  & outburst     &   $-$         &  (3)    \\
2450558.3446  &43140   &-0.00011   &0.00010  & outburst     &   $-$         &  (3)    \\
2450560.6630  &43167   &0.00027    &0.00010  & outburst     &   $-$         &  (3)    \\
2450560.7491  &43168   &0.00052    &0.00010  & outburst     &   $-$         &  (3)    \\
2450561.6935  &43179   &0.00054    &0.00010  & outburst     &   $-$         &  (3)    \\
2450561.7798  &43180   &0.00099    &0.00010  & outburst     &   $-$         &  (3)    \\
2450562.3794  &43187   &-0.00040   &0.00020  & outburst     &   $-$         &  (2)  \\
2450562.4654  &43188   &-0.00020   &0.00020  & outburst     &   $-$         &  (2)    \\
2450562.6369  &43190   &-0.00044   &0.00010  & outburst     &   $-$         &  (3)    \\
2450562.7227  &43191   &-0.00049   &0.00010  & outburst     &   $-$         &  (3)    \\
2450563.0667  &43195   &0.00010    &0.00020  & outburst     &   $-$         &  (2)    \\
2450563.6678  &43202   &0.00023    &0.00010  & outburst     &   $-$         &  (3)    \\
2450564.0123  &43206   &0.00130    &0.00020  & outburst     &   $-$         &  (2)    \\
2450564.0967  &43207   &-0.00010   &0.00020  & outburst     &   $-$         &  (2)    \\
2450564.6984  &43214   &0.00060    &0.00010  & outburst     &   $-$         &  (3)    \\
2450565.6430  &43225   &0.00082    &0.00010  & outburst     &   $-$         &  (3)    \\
2450565.7284  &43226   &0.00037    &0.00010  & outburst     &   $-$         &  (3)    \\
2450565.8145  &43227   &0.00062    &0.00010  & quiescence   &   $-$         &  (3)    \\
2450566.6731  &43237   &0.00069    &0.00010  & quiescence   &   $-$         &  (3)    \\
2450566.8448  &43239   &0.00068    &0.00010  & quiescence   &   $-$         &  (3)    \\
2450567.7888  &43250   &0.00031    &0.00010  & quiescence   &   $-$         &  (3)    \\
2450949.6625  &47698   &0.00156    &0.00010  & quiescence   &   $-$         &  (3)    \\
2451164.8951  &50205   &0.00165    &0.00010  & quiescence   &   $-$         &  (3)    \\
2451211.8562  &50752   &0.00137    &0.00010  & quiescence   &   $-$         &  (3)    \\
2451212.8861  &50764   &0.00104    &0.00010  & quiescence   &   $-$         &  (3)    \\
2451523.4147  &54381   &0.00072    &0.00010  & outburst     &   $-$         &  (3)    \\
2451523.5007  &54382   &0.00087    &0.00010  & outburst     &   $-$         &  (3)    \\
2451523.5862  &54383   &0.00052    &0.00010  & outburst     &   $-$         &  (3)    \\
2451523.9299  &54387   &0.00081    &0.00010  & outburst     &   $-$         &  (3)    \\
2451524.0160  &54388   &0.00106    &0.00010  & outburst     &   $-$         &  (3)    \\
2451526.5051  &54417   &0.00043    &0.00010  & outburst     &   $-$         &  (3)    \\
2451526.8496  &54421   &0.00152    &0.00010  & outburst     &   $-$         &  (3)    \\
2451526.9350  &54422   &0.00107    &0.00010  & outburst     &   $-$         &  (3)    \\
2451531.4854  &54475   &0.00128    &0.00010  & outburst     &   $-$         &  (3)    \\
\noalign{\smallskip}\hline
\end{tabular}
\end{center}
\end{table*}

\addtocounter{table}{-1}
\begin{table*}
\caption{$-$continued.}
 \begin{center}
 \small
 \begin{tabular}{ccccccc}\hline\hline
Min.(HJD)     &E       &O-C        &Errors   & State       &Telescopes     & Ref.$^a$    \\\hline
2451531.7424  &54478   &0.00072    &0.00010  & outburst     &   $-$         &  (3)    \\
2451531.8282  &54479   &0.00067    &0.00010  & outburst     &   $-$         &  (3)    \\
2451531.9140  &54480   &0.00061    &0.00010  & outburst     &   $-$         &  (3)    \\
2451532.6008  &54488   &0.00059    &0.00010  & outburst     &   $-$         &  (3)    \\
2451532.6867  &54489   &0.00064    &0.00010  & outburst     &   $-$         &  (3)    \\
2451578.9620  &55028   &0.00138    &0.00010  & quiescence   &   $-$         &  (3)    \\
2451579.0480  &55029   &0.00153    &0.00010  & quiescence   &   $-$         &  (3)    \\
2451584.7144  &55095   &0.00166    &0.00010  & quiescence   &   $-$         &  (3)    \\
2451584.7999  &55096   &0.00130    &0.00010  & quiescence   &   $-$         &  (3)    \\
2452780.46923 &69023   &0.00123    &0.00004  & quiescence   &   $-$         &  (4)    \\
2452782.44381 &69046   &0.00120    &0.00004  & quiescence   &   $-$         &  (4)    \\
2452783.47405 &69058   &0.00121    &0.00004  & quiescence   &   $-$         &  (4)    \\
2455148.36951 &96604	 &0.00048    &0.00010  & quiescence   &   2.4m        &  (6)    \\
2455162.36340 &96767	 &0.00039    &0.00010  & quiescence   &   2.4m        &  (6)    \\
2455189.32165 &97081   &0.00092    &0.00050  & quiescence   &   $-$         &  (5)    \\
2455190.26572 &97092	 &0.00061    &0.00010  & quiescence   &   2.4m        &  (6)    \\
2455971.18127 &106188	 &0.00075    &0.00005  & quiescence   &   2.4m        &  (6)    \\
2456711.66042 &114813	 &0.00108    &0.00070  & quiescence   &   $-$         &  (5)    \\
2456712.77675 &114826	 &0.00133    &0.00070  & quiescence   &   $-$         &  (5)    \\
2456714.83747 &114850	 &0.00158    &0.00070  & quiescence   &   $-$         &  (5)    \\
2457465.27595 &123591	 &0.00234    &0.00010  & quiescence   &   2.4m        &  (6)    \\
2457466.30594 &123603	 &0.00210    &0.00010  & quiescence   &   2.4m        &  (6)    \\
2457477.12345 &123729	 &0.00218    &0.00010  & quiescence   &   85cm        &  (6)    \\
2457777.26444 &127225	 &0.00242    &0.00010  & quiescence   &   1.0m        &  (6)    \\
\noalign{\smallskip}\hline
\end{tabular}
\end{center}
\scriptsize{ \quad \quad \quad \quad \quad $^a$References: (1) Howell et al. (1988); (2) Nogami et al. (2001); (3) Patterson et al. (2000);
\\. \quad \quad \quad \quad (4) Feline et al. (2004); (5) AAVSO data; (6) Our observations.}
\end{table*}

\begin{figure}[!h]
\begin{center}
\includegraphics[width=16cm]{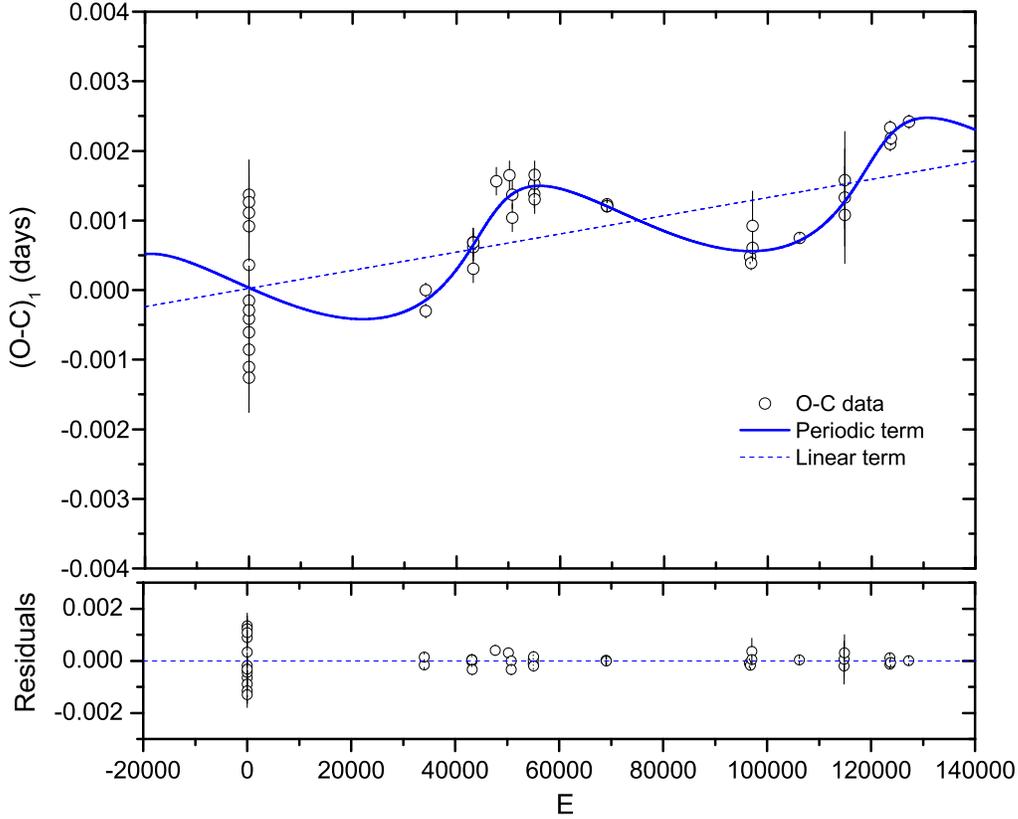}
\caption{$O-C$ diagram of DV UMa constructed with a linear plus LTT ephemeris. The black open circles refer to the data obtained in quiescence. The blue solid line in the upper panel represents the best-fitting model. The lower panel displays the fitting residuals from the complete ephemeris.}
\end{center}
\end{figure}

\begin{figure}[!h]
\begin{center}
\includegraphics[width=14cm]{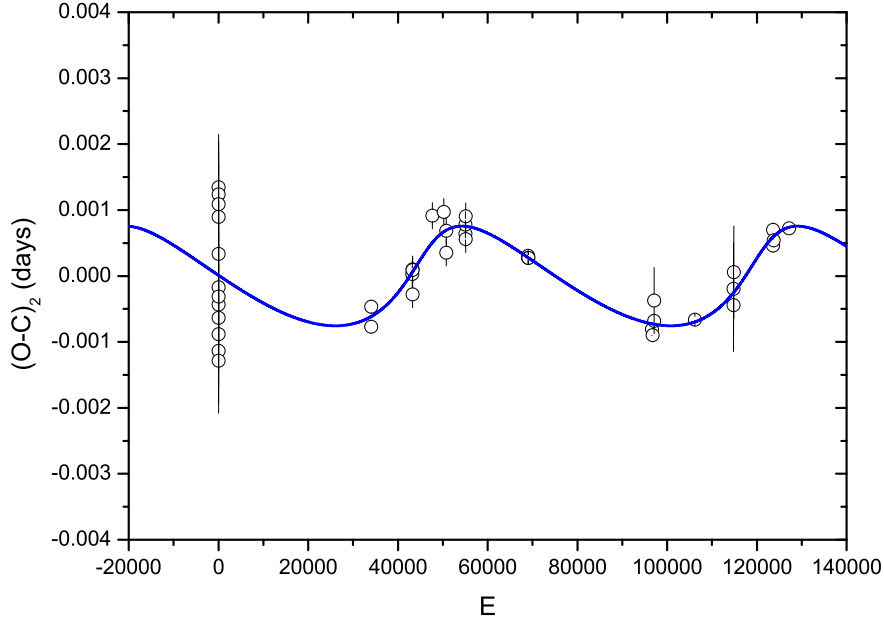}
\caption{The LTT orbit of a potential circumbinary companion extracted from the upper panel of Fig. 4. As shown, the cyclic oscillation can be seen clearly.}
\end{center}
\end{figure}

\begin{figure}[!h]
\begin{center}
\includegraphics[width=14cm]{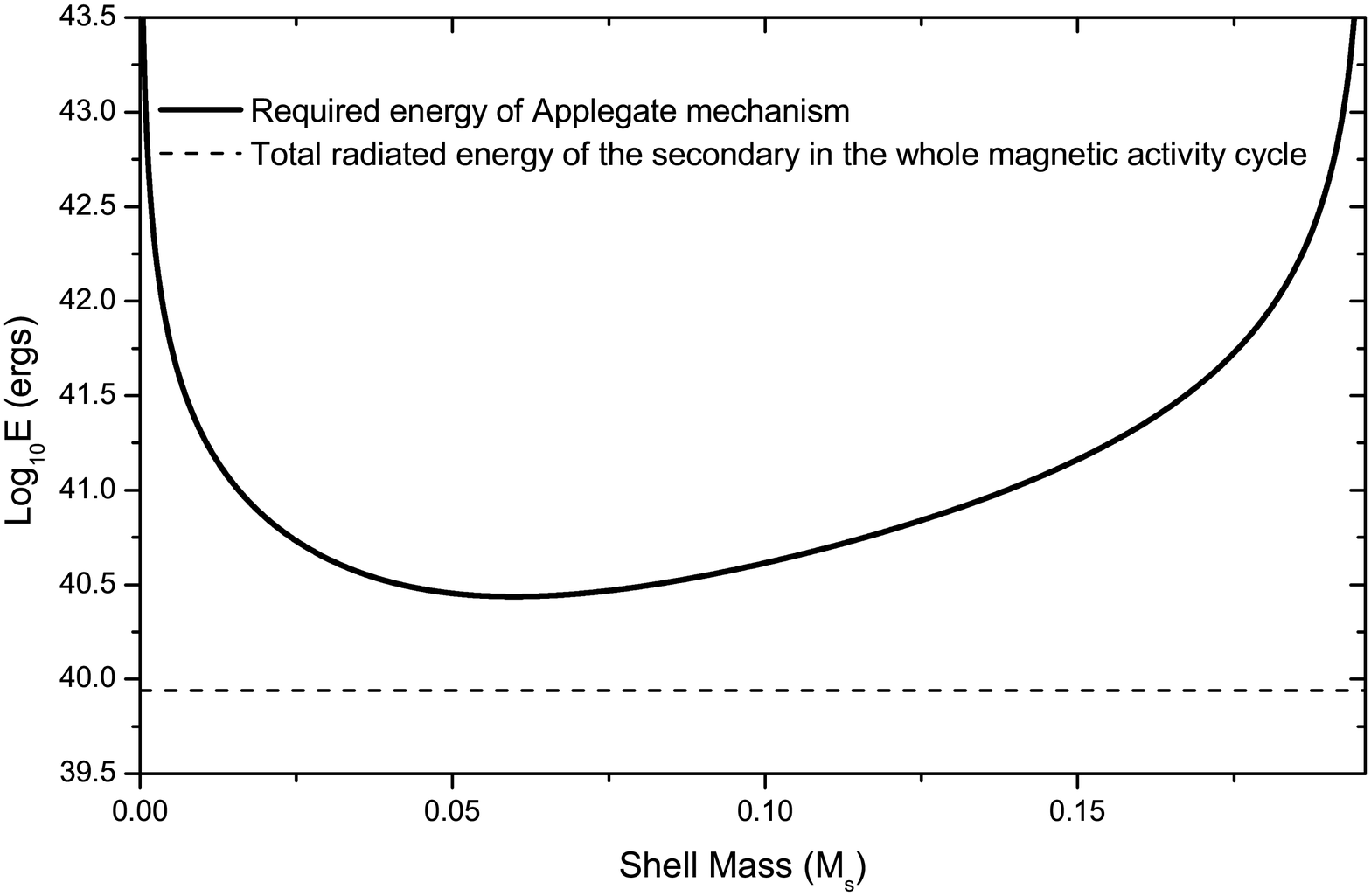}
\caption{The solid line shows the energy required to effect the observed cyclic change in DV UMa by using the Applegate mechanism. $M_s$ refers to the assumed shell mass of the secondary star. The dashed line represents the total energy radiates from the donor in a whole cycle.}
\end{center}
\end{figure}

\section{Eclipse timing variation and analysis}

The orbital period of DV UMa was investigated by previous authors but no sign of any period change was found.
Based on the discussion in Section 2, only the mid$-$eclipse times in quiescence are used for the present analysis. The updated $O-C$ diagram is shown in Fig. 4.
First of all, we represent the $O-C$ curve by a linear least-squares fit. However, the linear ephemeris cannot sufficiently represent all observed timings and the
residuals from it seem to show an apparent periodic oscillation. In general, the LTT via the presence of a circumbinary companion can be considered to be a possible cause of this oscillation (Irwin 1952). Therefore, the best-fitting model for $O-C$ diagram may be represented as
\begin{equation}
\begin{array}{lll}
(O-C)_1&=\Delta{T_{0}}+\Delta{P_{0}}{E}+K[(1-e^{2})\frac{\sin(\nu+\omega)}{1+e\cos\nu}+e\sin\omega] \\
&\\
&=\Delta{T_{0}}+\Delta{P_{0}}{E}+K[\sqrt{1-e^{2}}\sin{E^*}\cos\omega+\cos{E^*}\sin\omega],
\end{array}
\end{equation}
where $\Delta{T_{0}}$ and $\Delta{P_{0}}$ are the revised epoch and period. $\nu$ is the true anomaly, $E^*$ is the eccentric anomaly, $e$ is the orbital eccentricity and $K=a_{12}\sin{i^{'}}/c$ is the semi-amplitude of the LTT. To solve Equation (1), we used the two correlations:
\begin{equation}
N=E^{*}-e\sin{E^{*}},
\end{equation}
and
\begin{equation}
N=\frac{2\pi}{P_{3}}(t-T).
\end{equation}
$N$ is the mean anomaly and $t$ is the time of light minimum. In the process of fitting, the weighted least-squares method was used and the different weights were assigned to the different errors: the weights of 1, 5, 10 and 40 corresponding to the errors of 0.0005 d, 0.0002 d, 0.0001 d and 0.00005 d, respectively. All parameters and final results are summarized in Table 2.
The parameter errors are purely formal computed from the best-fitting covariance matrix.
The residual sum of squares (RSS) is $\sim0.00001$, indicating that the fit is very well.

Our result shows that the orbital period of DV UMa has a cyclic variation with an amplitude of $\sim71.1(\pm6.7)$ s and a period of $\sim17.58(\pm0.52)$ years. In the upper panel of Fig. 4, the dashed line denotes the linear-ephemeris revisions on the initial epoch and the orbital period.  the blue solid line represents the combination of the linear ephemeris and the periodic change. The lower panel shows the residuals from the best-fitting model. In order to display the periodic oscillation clearly, the $(O-C)_2$ values (by removing the linear ephemeris) are plotted in Fig. 5.

\begin{table}[!h]
\caption{Orbital parameters of the circumbinary brown dwarf companion.}\label{elements}
\begin{center}
\small
\begin{tabular}{lllllllll}
\hline
Parameters                                                          &Values                                \\
\hline
Revised epoch, ${\Delta{T_{0}}}$ (days)           & $+2.28(\pm1.46)\times10^{-5}$          \\
Revised period, ${\Delta{P_{0}}}$ (days)          & $+1.31(\pm0.16)\times10^{-8}$        \\
Eccentricity, $e$                                 & $0.44(\pm0.17)$  \\
Longitude of the periastron passage, $\omega$ (deg)     & $26.11(\pm21.55)$        \\
Periastron passage, $T$ (years)                          & $2444390.5(\pm424.5)$                      \\
The semi-amplitude, $K$ (days)                             & $0.000822(\pm0.000077)$  \\
Orbital period, $P_3$ (years)                                  & $17.58(\pm0.52)$                      \\
Projected semi-major axis, $a_{12}\sin{i^{'}}$ $(AU)$            & $0.143(\pm0.013)$  \\
Mass function,$f(m)$ $(M_{\odot})$                                  & $9.36(\pm2.61)\times10^{-6}$         \\
Mass of the third companion, $M_{3}\sin{i'}$ $(M_{\odot})$           & $0.025(\pm0.004)$                        \\
Orbital separation, $d_{3}$ ($AU,{i'=90^{\circ}}$)       & $8.6(\pm1.6)$                        \\
\hline
\end{tabular}
\end{center}
\end{table}

\section{Discussions and conclusions}

The main result in this paper is that the periodic variation in the orbital period of DV UMa is first detected and analyzed.
To explain the cyclic period oscillations in close binaries containing at least one cool star, one of the plausible causes is
a solar-type magnetic activity cycle in the late-type component (Applegate 1992). However, the Applegate mechanism may not work here because the donor in DV UMa is a fully convective main sequence star, its mass $M_2=0.196(\pm0.005)M_{\odot}$ (Savoury et al. 2011).
Moreover, the required energies to produce the observed change were computed using the same method proposed by Brinkworth et al. (2006). The result indicates that the required minimum energy is much larger than the total energy radiated from the secondary star in a whole cycle (see Fig. 6). Combining the parameters ($M_1=1.098M_{\odot}$, $M_2=0.196M_{\odot}$) presented by Savoury et al. (2011) with Kepler's third law we derived the orbital separation as $a=0.89R_{\odot}$.
Applying $T_{2}=3170K$ for the donor of $\sim0.196M_{\odot}$,
its luminosity can be estimated as $L_{2}=(\frac{R_{2}}{R_{\odot}})^{2}(\frac{T_{2}}{T_{\odot}})^{4}L_{\odot}$.
Further, the Applegate mechanism has been studied in depth by V\"{o}lschow et al. (2016) and an improved Applegate model also has been explored. They concluded that an ideal Applegate post-common-envelope binary (PCEB) should consist of a very close orbit ($\sim0.5R_{\odot}$) and a secondary star of $\sim0.5M_{\odot}$. The Applegate mechanism is more feasible for a much larger donor mass than is present here.
Clearly, DV UMa can not satisfy any of those conditions.
Therefore, we suggest that the Applegate mechanism may not be responsible for this cyclic change. The most likely interpretation of the period oscillation is the LTT via the presence of an unseen companion. But this conclusion dose not deny the presence of magnetic activity in the secondary star. On the contrary, this only implies that the magnetic activity is not the dominant mechanism here because it can not provide enough energy to produce the observed period variation.

The parameters of the best-fitting model (see Table 2), coupled with the absolute parameters determined by Savoury et al. 2011,  lead to the mass function and the mass of the tertiary companion as
$f(m)=9.36(\pm2.61)\times10^{-6}M_{\odot}$ and $M_{3}\sin{i'}=0.025(\pm0.004)M_{\odot}$. Assuming a random distribution
of orbital plane inclinations, when $i'\geq 20.9^{\circ}$, the mass corresponds to $0.025\leq M_{3} \leq 0.07 M_{\odot}$, which would match to a brown dwarf.
However, the circumbinary companions are expected to be coplanar to the orbital plane of the eclipsing pair (Bonnell \& Bate 1994). So, if the third body orbiting DV UMa is on a coplanar orbit (i.e. $i'=i=82.9^{\circ}$), the mass is derived as $M_{3}=0.0252M_{\odot}$, it would be a brown dwarf. The orbital radius $d_3$ of the tertiary companion is $\sim8.6(\pm1.6)$ AU (when $i'=90^{\circ}$).

So far, only a few brown dwarfs orbiting single white dwarfs were discovered by several previous authors (e.g. Becklin \& Zuckerman 1988; Farihi \& Christopher 2004; Dobbie et al. 2005; Burleigh et al. 2006; Maxted et al. 2006). But in comparison with single white dwarfs, the brown dwarfs around the white dwarf binaries are more rare. Recently, only two white dwarf binaries were reported containing the brown dwarf companions, i.e. V471 Tau (Guinan \& Ribas 2001) and SDSS J143547 (Qian et al. 2016). Also note that the brown dwarf companion in V471 Tau was excluded by Hardy et al. (2015) but \textbf{a} new study agrees with the presence of a brown dwarf companion (Vaccaro et al. 2015).
Moreover, some possible planets orbiting the eclipsing CVs or pre-CVs were also presented such as V893 Sco (Bruch 2014), V2051 Oph (Qian et al. 2015), OY Car (Han et al. 2015), DP Leo (Qian et al. 2010; Beuermann et al. 2011), HU Aqr (Qian et al. 2011; Go\'{z}dziewski et al. 2015), UZ For (Dai et al. 2010; Potter et al. 2011) and NN Ser (Marsh et al. 2014). These discoveries suggest that the PCEBs may be one of the most common host stars.
The origins of substellar objects are very complex: they may have existed before the common-envelope (CE) event (first generation) (e.g. Qian et al. 2016), or they may be second generation planets (or brown dwarfs) formed during the CE phase (e.g. V\"{o}lschow et al. 2014). If they are the first generation companions, their masses might be increased by accreting a large amount of material during the CE stage, and more massive planets could become brown dwarf companions. But it is difficult to distinguish between the two types of substellar populations at present.
The discovery of a potential brown dwarf orbiting DV UMa may shed new light on our understanding of the origin and the evolution of the circumbinary substellar objects.
As yet, however, the data coverage on DV UMa is less than two cycles of the cyclic change. Further observations are critically required to ascertain whether the observed oscillation is strictly periodic, or only quasi-periodic.

\acknowledgments{This work is supported by the Chinese Natural Science Foundation (Grant Nos. 11325315, 11611530685, 11573063 and 11133007), the Strategic Priority Research Program ''The Emergence of Cosmological Structure'' of the Chinese Academy of Sciences (Grant Nos. XDB09010202) and the Science Foundation of Yunnan Province (Grant Nos. 2012HC011).
This study is supported by the Russian Foundation for Basic Research (project Nos. 17-52-53200).
New CCD photometric observations of DV UMa were obtained with the 1m and the 2.4m telescopes at the Yunnan Observatories, the 85cm telescope in Xinglong Observation base in China.
We thank the numerous observers worldwide who contributed many observations of DV UMa to the AAVSO database. Finally, we thank the referee for an insightful report that has significantly improved the paper.}

\end{document}